\documentstyle[aps,prb,epsf]{revtex}
\begin{document}

\twocolumn[\hsize\textwidth\columnwidth\hsize\csname
@twocolumnfalse\endcsname

\title{Fragile-strong transitions and polyamorphism in glass former fluids}
\author{E. A. Jagla}
\address{Centro At\'omico Bariloche, Comisi\'on Nacional de Energ\'{\i}a
At\'omica \\(8400) S. C. de Bariloche, Argentina}
\maketitle
 
\begin{abstract}
A simple model of a glass former fluid, consisting of a bidisperse mixture
of penetrable spheres is studied. The model shows a transition 
from fragile to strong behavior 
as temperature is reduced.
This transition is driven by the competition between the two mechanisms that contribute
to diffusivity in the model:
collective rearrangement of particles (responsible for the fragile behavior), and 
individual particle motion (which gives rise to the strong behavior at
low temperature).
We also observe a maximum of diffusivity as a function of
pressure that can be interpreted within the same framework.
The connection between this behavior and polyamorphism is addressed.
\end{abstract}
 
\vskip2pc] \narrowtext

\section{introduction}

The dependence of single particle diffusivity $D$ (or alternatively 
viscosity $\eta$)  on temperature  is the main indicator that allows to classify glass formers as 
``fragile'' or ``strong''\cite{angell}.
For strong glass formers (such as SiO$_2$) $D(T)$ is well fitted by an Arrhenius plot,
indicating that there is a well defined value of the energy barriers which are 
relevant to diffusion.
For fragile glass formers, diffusion is the result of collective rearrangements of the
position of particles. As temperature is reduced, 
the size of the correlated clusters increases, and $D$ decreases more rapidly
than the Arrhenius form\cite{agdm}. 
There is evidence that some materials show a fragile
behavior at high temperatures,
whereas $D(T)$  crosses over to a strong behavior at low temperatures. Water is an
example\cite{waterfs}. 
The microscopic mechanism that originates this behavior is not well
understood.

Some materials display the phenomenon of polyamorphism, namely, 
the existence of different amorphous (solid) configurations at very low temperature. 
One well known
example is, again, water\cite{mishima}. 
In this case the two different amorphous states
transform into each other through a first order transition
when pressure is applied at low temperatures.
Polyamorphism has been numerically found in
a model for SiO$_2$ \cite{sio2pa}
and in some tetrahedrally 
coordinated materials: Si, Al$_2$O$_3$-Y$_2$O$_3$, and others \cite{ms}.

One route to the appearance of polyamorphism is the possibility of two different local 
arrangement of particles, of different density. If this is the case (and this will depend on the 
interaction potential between particles), external pressure can
make the structure collapse from the most expanded configuration to the densest one, 
near some typical pressure
$P_{cr}$. It has been recently shown in a simple model of spherical, purely repulsive particles 
with these characteristics \cite{jcp2}, that
the dependence of density on pressure 
at very low temperatures has an anomalously rapid increase (in other words, that the isothermal
compressibility has a maximum) for pressures around some
crossover value $P_{cr}$ at which the two local environments become equally stable.
We will call this a `weak' form of polyamorphism.
If some kind of attraction between particles is added, this weak polyamorphism
may transform (depending on the details of the attraction)
in a true polyamorphic first order transition between two different amorphous 
structures\cite{jcp2}. 
The interaction potential used in [7] to obtain two different local 
environments consists of a hard core at distance $r_0$ and a shoulder that extends up to 
some other distance $r_1>r_0$. Interparticle distances $r\sim r_0$ are favored at high $P$, whereas
at low $P$,  $r\sim r_1$ is preferred. 

There is an intimate relation between (weak) polyamorphism
and fragile-strong transitions. We will show in this paper that a model 
within the class of those discussed in Ref. [7], displays (weak) polyamorphism
and has a transition from fragile to strong behavior when $T$
is reduced. The model also exhibits a maximum in $D$ as a function of pressure, which 
is well known to
occur, for instance, in water\cite{aguamax}. All these phenomena are 
consequences of the existence of two different local environments available for the particles, 
and the two qualitatively different mechanisms that contribute
to the diffusivity in these systems, as explained below.

\section{The model}

We studied a model similar to that of Ref [7], with hard core radius 
$r_0=0$. In this way, particles are {\it
penetrable}, and can overlap.
A bidisperse system is considered, to avoid crystallization.
Particle $i$ is characterized by the value of $r_1^i$, which is taken from a bimodal distribution,
i.e., $r_1^i=r_a$, or $r_1^i=r_b$, with equal probability. The interaction potential between
particles $i$ and $j$ takes the form

\begin{eqnarray}
V^{ij}(r)&=&\varepsilon_0 (1-r/R)(R/\bar r)^3~~~~{\rm for}~~~~r<R\nonumber\\
V^{ij}(r)&=&0 ~~~~{\rm for}~~~~~ r>R  \label{vder}
\end{eqnarray}
where $R=r_1^i+r_1^j$, $\bar r$ is the average radius of the particles,
and $\varepsilon_0$ sets the energy scale\cite{nota}. In all the results to be presented, we
use $r_a=0.4375 \bar r$, and $r_b=0.5625 \bar r$.
Temperature will be given in units of $\varepsilon_0/k_B$, and pressure in units of 
$\varepsilon_0/\bar r ^3$.

We used standard Monte Carlo techniques, in the NPT ensemble. 
The system is given the dynamics
of the Monte Carlo evolution, and diffusivities are calculated using this as the 
temporal evolution.
At each time step the position of a single particle is modified to a new
position which is randomly chosen within a sphere of radius 0.06 $\bar r$ centered at the
original position. This trial movement is accepted according to the Metropolis rule. The update is
made sequentially for all particles. One Monte Carlo step corresponds to a sweep over all
particles.

Previous studies of related potentials \cite{jcp2,pre,sadr} have shown that they display
some `anomalous' behavior, 
such as a line of maximum density and compressibility in the $P$-$T$ diagram, and regions of 
anomalous melting in the $P$-$T$ plane. These anomalies match those of real materials (such as water,
SiO$_2$, etc.) where 
polyamorphism and fragile-strong transitions are observed or expected.
We will concentrate here in
the fluid phases of the model (see [10] for a discussion of the crystalline
phases which are non-trivial) focusing on the behavior of diffusivity $D$.
This is calculated from the long time behavior of the distance traveled by the particles, 
$<({\bf r}(t\rightarrow \infty)-{\bf r}(0))^2>= 6Dt$, where 
${\bf r}(t)$ is the position of the particle at time $t$. The reported values always correspond to an
average over all particles in the system. $D$ will be given in units of $\bar r^2/{t_0}$,
where $t_0$ is the time corresponding to one Monte Carlo step. 
The number of Monte Carlo steps which are necessary
to get good statistics greatly depends on the parameters $P$ and $T$. As much as $5\times 10^7$ steps
were simulated for the lowest temperatures reported. The results correspond to a system of
200 particles, with periodic boundary conditions.

\section{Results}

We start showing results indicating the existence of weak polyamorphism in the model. In Fig.
\ref{f1} we see
the evolution of the density $\rho$ of the system and the corresponding values of
isothermal compressibility $K_T\equiv \rho ^{-1}\partial \rho/\partial P $ at $T=0$ and $T=0.03$. 
Each point was
obtained by an independent simulation at fixed pressure, and reducing temperature.
There is a rather abrupt raising in density and a maximum in $K_T$ around $P\sim 1.3$ at $T=0$. The
maximum becomes smoother and its position 
moves to slightly higher pressures at higher temperatures. The existence of this  maximum 
is easy to understand. 
At low pressures (and low temperatures) the system behaves as a set of hard spheres, overlapping of
particles in the same spatial position is highly improbable due to energetic reasons. Upon increasing
external pressure, a value $P_{cr}$ is reached beyond which it is energetically more favorable 
to accommodate particles in pairs, at each spatial position. The change of stability between
`singled' and `doubled' local configurations is
responsible for the maximum in $K_T(P)$ around $P_{cr} = 1.3$ \cite{nota0}.
This is consistent with the form of the radial distribution function $S(r)$, plotted in Fig.
\ref{f2} for $T=0.03$, at different values of $P$, where we see
the rapid increase of weight around $r\sim 0 $ (indicating the appearance of `doubled' particles) 
when crossing $P_{cr}$. This is the evidence for weak
polyamorphism in the model.

\begin{figure}
\narrowtext
\epsfxsize=3.3truein
\vbox{\hskip 0.05truein
\epsffile{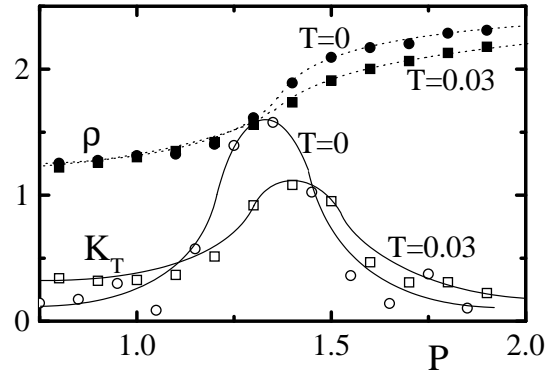}}
\medskip
\caption{Pressure dependence of the density $\rho$ and isothermal compressibility $K_T$, 
at $T=0$ and $T=0.03$. Lines are guides to the eye.}
\label{f1}
\end{figure}

\begin{figure}
\narrowtext
\epsfxsize=3.3truein
\vbox{\hskip 0.05truein
\epsffile{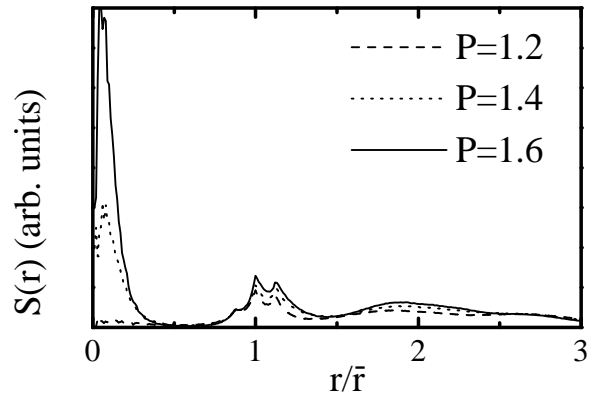}}
\medskip
\caption{The radial distribution function at $T=0.03$, for three different values of $P$.
There is a rapid increase of weight at distance
$r\sim r_0=0$ for $P\sim P_{cr}$. The three peaks around $r=\bar r$ originate in the 
bidispersity of the system.}
\label{f2}
\end{figure}

Next, we show results for single particle diffusivity as a function of pressure $D(P)$, at different 
values of $T$ in Fig. \ref{f3}. We see clearly a maximum in $D$ at $P\sim P_{cr}$. 
The origin of this maximum can be 
understood in the following way. For pressures much lower (larger) than the crossover value $P_{cr}$
mostly singled (doubled) configuration of particles exist \cite{nota2}. 
For $P \sim P_{cr}$, both singled and doubled particles coexist in the sample. Moreover, local
rearrangements in which one particle of a doubled pair moves onto some nearby singled particle are
possible. This diffusion mechanism (which is illustrated in Fig. \ref{f4}) is available 
only for $P\sim
P_{cr}$, and produces an enhancement of diffusivity in this pressure range.

\begin{figure}
\narrowtext
\epsfxsize=3.3truein
\vbox{\hskip 0.05truein
\epsffile{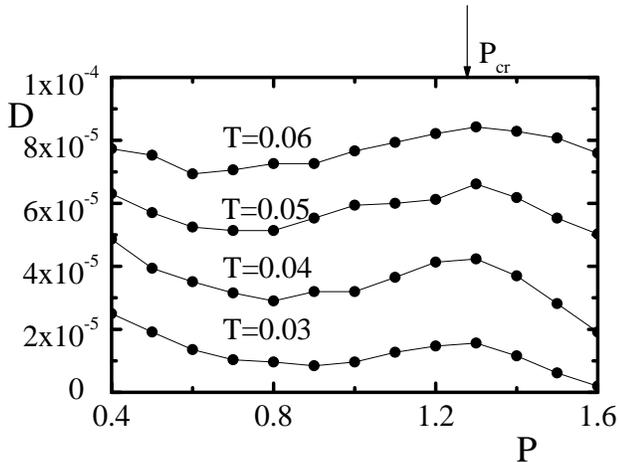}}
\medskip
\caption{Diffusivity as a function of pressure, at different temperatures. An anomalous maximum
is clearly visible around $P_{cr}$.}
\label{f3}
\end{figure}

\begin{figure}
\narrowtext
\epsfxsize=3.3truein
\vbox{\hskip 0.05truein
\epsffile{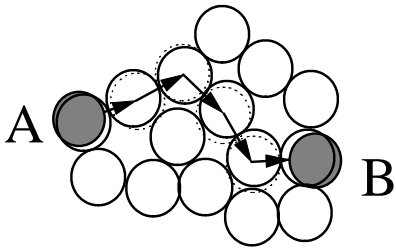}}
\medskip
\caption{Two-dimensional sketch of the mechanism responsible for additional diffusivity when `singled'
and `doubled' particles coexist in the system. One of the particles in a pair (the gray one) 
moves from $A$ to $B$
by successively jumping onto neighbor singled particles. Notice 
that this mechanism is not a collective one.}
\label{f4}
\end{figure}

The third set of results concerns the temperature dependence of diffusivity. 
In a standard glass former
(i.e., pure hard spheres, or Lennard-Jones particles) diffusivity occurs {\it via} rearrangements
of clusters of particles. As temperature is reduced, the size of the typical cluster that can rearrange
increases, producing the typical temperature dependence of diffusivity of fragile glass formers. This
mechanism is present in our system, and is basically the only one active for $P$ much larger or lower than
$P_{cr}$. But for $P\sim P_{cr}$ the mechanism discussed in the previous paragraph is also available.
Since this is essentially a single particle mechanism (with a well defined activation energy), it 
account for a contribution to diffusivity which is basically Arrhenius-like. 
The total diffusivity is the
combination of the `fragile' part plus this `strong' part\cite{jpcm}. At large temperature the fragile part
dominates. On reducing temperature the fragile contribution decreases much more rapidly, and leaves the
strong contribution to be the main part at very low temperatures. 
In Fig. \ref{f5} we see the results for $D(T)$ in our system, at three different values of
$P$. There is, in fact, a crossover between
a fragile behavior at high temperature and a strong behavior at low temperature. The fragile
part of these curves can be fitted by a Vogel-Fulcher law\cite{vf,angell} ($D\sim D_0 \exp(A/(T-T_0)$)
for the diffusivity of fragile glass formers, but the whole curve cannot be fitted to a single
expression of this type, as we see in the figure.
Also to be
noticed is the fact that the strong component of diffusivity at low temperature is maximized
around $P_{cr}$. So it is close to $P_{cr}$ than the fragile strong transition is more easily 
observed.
To compare with a more usual case, we made an independent
simulation for particles with a strict hard core at $R\equiv r_1^i+r_1^j$. In this case the
properties of the system depend only on $P/T$ (or alternatively, on density), i.e, the system 
is athermal.
But we still plot $D$ vs $T^{-1}$ at a fixed $P=1.2$ to compare with the general case. 
The results are the symbols at the left of Fig. \ref{f5}. In this case the whole curve can be
nicely fitted by a Vogel-Fulcher law, 
the fragile behavior extending 
down to the lowest values of diffusivity that we can detect.

\begin{figure}
\narrowtext
\epsfxsize=3.3truein
\vbox{\hskip 0.05truein
\epsffile{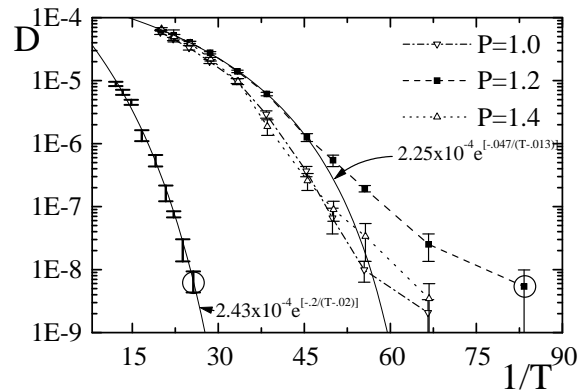}}
\medskip
\caption{Diffusivity as a function of temperature for different pressures around $P_{cr}$. 
The high temperature part of the results 
can be fitted by a Vogel-Fulcher law, as shown for the $P=1.2$ case. 
At low temperature diffusivity crosses over to a new regime,
corresponding to a strong behavior. Left curve: results of a simulation for 
particles with a strict hard core at distance $R$, for $P=1.2$. Now 
the whole behavior fits in a Volger-Fulcher law, as indicated. The circles indicates the points at which
the snapshots of Fig. [6] were taken.}
\label{f5}
\end{figure}

To get a visualization of the two different mechanisms for diffusivity we present the snapshots 
of Fig. \ref{f6}. They show the projection of the positions of the particles onto a reference
plane, and the displacements of the particles during a simulation of
$5\times 10^7$ Monte Carlo steps. Panel (a) was obtained at $P=1.2$, $T=0.012$, in our system 
of particles interacting through (\ref{vder}). 
According to Fig.
\ref{f5}, diffusivity has a strong behavior in this case. Panel (b) was obtained for the
system with strict hard cores at distance $R$, at $P=1.2$, $T=0.039$. 
In this case diffusivity has a fragile behavior. Note that the
overall values of $D$ in both cases (a) and (b) are comparable (see Fig. \ref{f5}). However, the
observable features of diffusivity are clearly different. In the fragile case (Fig.
\ref{f6}(b)) a cluster of particles (approximately 
delimited by the dotted line, note however that this is only qualitative, since the system is 
three-dimensional) has rearranged during the simulation time. This is
consistent
with the collective nature of the rearrangements in the fragile regime. For the
strong case however, as Fig. \ref{f6}(a) shows, we see that mostly independent motions of particles in
uncorrelated positions of the sample have occur.
This is the manifestation of the single particle character of diffusion in this case, and is
consistent with the pictorial image of Fig. \ref{f4}.

\begin{figure}
\narrowtext
\epsfxsize=3.3truein
\vbox{\hskip 0.05truein
\epsffile{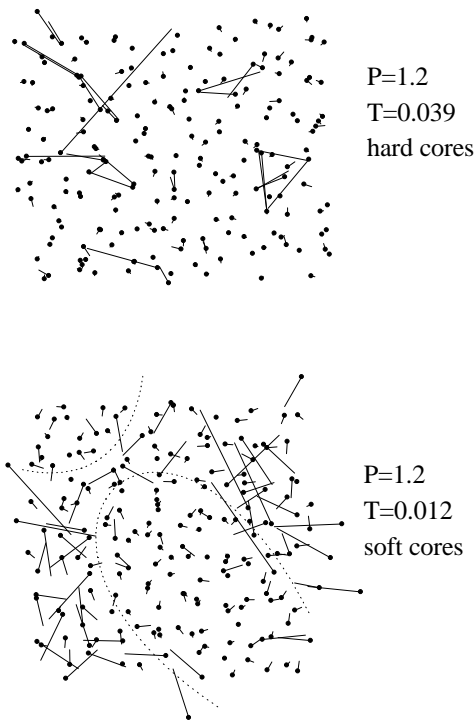}}
\medskip
\caption{Projections of the displacements of the particles during a simulation of 10$^7$ Monte Carlo
steps, at values of
$P$ and $T$ as indicated. In (a) particles are taken as strict hard spheres of radius $r_1$,
whereas in (b) the interaction potential is that of (\ref{vder}). 
The diffusivity corresponds to that of a strong material in (a), and a fragile 
material in (b).}
\label{f6}
\end{figure}


\section{Discussion and Conclusions}

We have studied the relation between
fragile-strong transitions and polyamorphism in a model glass former.
We showed that if the system has a 
tendency to reach two different amorphous configurations when cooled down below 
and above some crossover pressure $P_{cr}$, then on general grounds we can expect 
maxima in the diffusivity $D$ as a function of $P$, occurring precisely around 
$P_{cr}$. In addition, $D(T)$ calculated at $P\sim P_{cr}$ is expected to have 
a fragile-strong transition when temperature is reduced. 

This behavior is in fact expected in the wide class of tetrahedrally coordinated materials.
In all these materials, there are
two typical local arrangement of particles, one more expanded (corresponding for 
instance to the local structure of ice Ih in the case of water) 
and another more compact (corresponding to local 
arrangements in high pressure polymorphs of ice), which correlate with our two typical 
distances between particles in the model. Then we expect in all these cases fragile-strong
transitions and maxima of $D(P)$ to be present.

Real materials include in general some sort of attraction between particles. With this ingredient
weak polyamorphism may develop into a true first order phase transition between two different 
amorphous phases. This is in fact what happens in water, which is the best 
known example in which a truly first order polyamorphic transition, a 
fragile-strong transition and maxima in $D(P)$ occur. In this case the two 
amorphous structures are separated by a first order line that is supposed to end in a critical point 
in the supercooled liquid region. The effects we have discussed (maxima in $D(P)$ and 
fragile-strong transitions) however do not require 
the existence of this critical point. Only weak polyamorphism is needed.

\section{Acknowledgements}

This work was financially supported by Consejo Nacional de Investigaciones Cient\'{\i}ficas
y T\'ecnicas (CONICET), Argentina.


\begin{references}
\bibitem{nota}With this interaction potential, two particles at the same position get
an energy $\varepsilon _0 (R/\bar r)^3$. Since the energy per particle associated to 
the existence of an external
pressure $P$ is $Pv \sim PR^3$, our choice is reasonable in order to 
obtain a single crossover pressure $P_{cr}$ (independent of the radius of the particle) at
which particles overlap.
\bibitem{nota0} For a crystalline material, in which all particles are in equivalent positions
the transition from singled to doubled configurations occurs at a well defined pressure (at $T=0$). For
our glassy system in which there are many inequivalent spatial positions, this transition is 
smooth, even at $T=0$, as Fig. \ref{f1} shows.
\bibitem{nota2} At sufficiently high pressures, particles also overlap in triplets, etc. In
the range of pressures we are working these configurations are extremely rare.
\end{references}
\end{document}